# A Precipice Below Which Lies Absurdity?

## Theories without a Spacetime and Scientific Understanding

Sebastian De Haro[1] and Henk W. de Regt[1]

[1]Trinity College, Cambridge, CB2 1TQ, United Kingdom
Department of History and Philosophy of Science, University of Cambridge, United Kingdom
Vossius Center for History of Humanities and Sciences, University of Amsterdam, The Netherlands
[2]Faculty of Philosophy, Vrije Universiteit Amsterdam
De Boelelaan 1105, 1081 HV Amsterdam, The Netherlands

sd696@cam.ac.uk,    h.w.de.regt@vu.nl

7 July 2018[1]

**Abstract**

While the relation between visualization and scientific understanding has been a topic of long-standing discussion, recent developments in physics have pushed the boundaries of this debate to new and still unexplored realms. For it is claimed that, in certain theories of quantum gravity, spacetime 'disappears': and this suggests that one may have sensible physical theories in which spacetime is completely absent. This makes the philosophical question whether such theories are intelligible, even more pressing. And if such theories are intelligible, the question then is how they manage to do so. In this paper, we adapt the contextual theory of scientific understanding, developed by one of us, to fit the novel challenges posed by physical theories without spacetime. We construe understanding as a matter of skill rather than just knowledge. The appeal is thus to *understanding*, rather than *explanation,* because we will be concerned with the tools that scientists have at their disposal for understanding these theories. Our central thesis is that such physical theories can provide scientific understanding, and that such understanding does not require spacetimes of any sort. Our argument consists of four consecutive steps: (a) We argue, from the general theory of scientific understanding, that although visualization is an oft-used tool for understanding, it is not a necessary condition for it; (b) we criticise certain metaphysical preconceptions which can stand in the way of recognising how intelligibility without spacetime can be had; (c) we catalogue tools for rendering theories without a spacetime intelligible; and (d) we give examples of cases in which understanding is attained without a spacetime, and explain what kind of understanding these examples provide.

*Keywords:* scientific understanding, intelligibility of theories, quantum gravity, emergence of spacetime

---

[1] Forthcoming in *Synthese.*

# Contents

## 1. Introduction

The debate over the indispensability of representing physical systems in space and in time is an old one. Descartes already took extension to be one of the defining properties of matter, sharply contrasting the properties of the thinking substance, which he regarded as purely abstract. Kant turned space and time into 'forms of intuition' (*Anschauungsformen*), thus stressing even more the necessity of the spatial and temporal representations for our grasping of the phenomena. This was not merely a philosophical debate divorced from scientific issues: late nineteenth- and early twentieth-century scientific developments, from Poincaré's conventionalism and Einstein's two theories of relativity to quantum mechanics, challenged the view that space and time—whether a priori concepts of the mind, or some kind of ultimate foundation of reality—are entities which are fixed once and for all. On the other hand, a physicist like Schrödinger, who worked at the forefront of the modern revolution in physics, used the idea of the necessity of spatial and temporal representations for our grasping of the phenomena to argue for the superiority of his own explicitly spatiotemporal wave mechanics —criticising the lack of *Anschaulichkeit* of Heisenberg's abstract matrix mechanics. However, with the advent of quantum field theory, where also the concept of a point particle as an object endowed with persistence in space and time becomes obsolete (since elementary particles are transmuted into each other through interactions, thereby losing their stability), the picture of a world that is straightforwardly visualizable in space and time was further eroded.

The most recent developments in physics add new perspectives to the debate about visualizability. Contemporary theories of physics suggest that space and time are perhaps not fundamental at all. They are thought to imply that, at the basic level, space and time cease to exist as fundamental entities (often, for some regime of the physical parameters yet inaccessible to experiment), in a way still more radical than envisaged by general relativity and quantum field theory.

It is important to appreciate the novelty of this claim and of the philosophical questions that the current debate spurs, compared to earlier discussions of the *Anschaulichkeit* of quantum mechanics. Quantum mechanics introduced uncertainties into the spatiotemporal properties of entities *moving in* space and time, but spacetime itself stayed firmly in place, as a stage on which these entities could exist, move and interact. Electrons, photons and atoms are all observed, *measured in* spacetime, with the help of detectors that record their properties in space and in time. On the other hand, the new theories of gravity which we are referring to here are theories of *quantum* gravity, which are concerned with the fabric of spacetime itself, and the claim often made is that this very fabric ceases to be a fundamental entity: there is 'no spacetime' at the fundamental level.

This prompts the philosophical question of whether such theories without a spacetime, which are inherently unvisualizable, can be intelligible at all, and whether they can provide us with understanding of the phenomena. There is a growing philosophical literature on scientific understanding.[2] One of the questions that this literature has addressed is how theories that resist direct visualization can lead to understanding.[3] On a pragmatic theory of scientific understanding, the answer to this question is straightforward: although direct visualization is an important tool for achieving understanding, it is not essential to it. For scientists can use indirect forms of visualization, and also other tools, to render their theories intelligible, and to understand the phenomena. This suggests that theories of scientific understanding can be fruitfully brought to bear on the question of how quantum gravity theories can be rendered intelligible. Thus, the following Basic Question suggests itself:

(BQ)   Can physical theories in which space and time are absent from the basic theory be rendered intelligible?

In this paper, we will address (BQ) by engaging with some of the recent literature on the merits of 'theories without a spacetime', i.e. theories with no spacetime at the basic level. In particular, we will articulate our approach to (BQ) using four further questions:

(i)   Which answers to (BQ) are suggested by recent philosophical research in the field of scientific understanding? Our working hypothesis is that the answer should be affirmative: theories without spacetime can indeed provide scientific understanding.

(ii)   Which arguments support the hypothesis that these theories provide understanding? This question will be discussed in the light of misgivings on a positive answer to (BQ) in the literature, as voiced by Maudlin (2002), and Lam and Esfeld (2013), whose arguments we will analyse.

---

[2] See Baumberger, Beisbart and Brun (2017) for an overview of the current debates on understanding in epistemology and philosophy of science.
[3] For an overview of the role of visualization in scientific understanding, and in particular for a discussion of whether visualization is indispensable for scientific understanding, see Mößner (2015).

(iii) Can recent claims and debates in the literature on the philosophy of spacetime be rephrased as definite answers to, or positions on, (BQ)? That is, can we construe these debates as disagreements about whether or not 'theories without a spacetime' are intelligible, and/or as disagreements about whether they can provide scientific understanding?

(iv) Does the literature on quantum gravity provide concrete examples of how physicists attain understanding without spacetime?

The literature on quantum gravity has so far emphasised questions such as for example emergence and empirical coherence,[4] but it has left the question of scientific understanding unanalysed. This is dissatisfying, for two reasons: First, scientific understanding is a philosophical topic in its own right; and, particularly in view of the discussion of visualization we reviewed above, it is pressing to ask how the further erosion of visualization in quantum gravity impacts understanding. Second, although some of the debates in the literature are about how to interpret theories without a spacetime in a straightforward physical sense, underlying these surface disagreements are distinct conceptions of scientific understanding, or so we will argue.

In Section 2, we take up point (i): we analyse (BQ), using the contextual theory of scientific understanding developed by one of us (HdR). We also address (ii) and dispel philosophical reasons one may have for being sceptical about understanding in theories without a spacetime. In Section 3, we take up point (iii) by reviewing two debates in the literature about the intelligibility of 'theories without a spacetime'. We focus on recent debates in the philosophy of physics literature rather than on the 'theories without a spacetime' themselves, as construed by the physicists (we will turn to the physics literature in Section 4). We will reframe the debates in terms of different views on (BQ). Section 4 takes up question (iv) by analysing explicit examples in the physics literature, and proposes a set of tools which are available for understanding in quantum gravity (these examples and tools are dealt with more fully in a companion paper, De Haro and De Regt (2018)).

## 2. Can 'theories without a spacetime' provide scientific understanding?

In the present section, we address questions (i) and (ii) listed in the introduction. In §2.1 we first review the current debate in philosophy of science and show how the contextual theory of scientific understanding, developed by one of us, illuminates developments in twentieth-century physics that challenged space-time description and the associated *Anschaulichkeit*. In §2.2 we will address a possible objection against our view: metaphysical apriorism.

### 2.1. What is scientific understanding and how can it be attained?

Only recently has understanding become a topic of investigation by philosophers of science. Scientific understanding is typically produced by explanations.[5] For many years, philosophers focused on the notion of explanation, because explanation was regarded as objective and susceptible to logical analysis, while understanding, by contrast, was assumed to be purely subjective and accordingly only of interest to psychologists and historians of science.[6] In the 1970s and 1980s, however, philosophers of science began to take seriously the relation between explanation and understanding. Some argued that explanations yield understanding by providing a *causal* account of how a phenomenon occurs, while others defended the idea that they do so by incorporating the phenomenon in a *unified* system. The deeper question of why causal and/or unifying accounts were capable of producing understanding was not addressed, however. It was only after the turn of the millennium that philosophers of science started to analyse the notion of (scientific) understanding in a systematic way. Today there is a lively philosophical debate on the nature of and conditions for scientific understanding, which features increasing interaction between philosophers of science and epistemologists.[7]

---

[4] For an overview, see the special issue edited by Huggett and Wüthrich (2013).
[5] While this is the standard view, there is currently a debate among philosophers of science about the question of whether there can be understanding without explanation; see De Regt (2013) and the papers in the special section of *Studies in History and Philosophy of Science* to which this is an introduction.
[6] See De Regt (2017, Chapters 2 and 3).

A central issue in the current debate on understanding is the question of how two seemingly rival conceptions of (scientific) understanding relate: understanding as a species of knowledge and understanding as a skill (or ability). The view that understanding is merely a specific type of knowledge is implicit in many traditional approaches in epistemology and philosophy of science. The above-mentioned theories of understanding are examples: they assume that understanding is knowledge of causal relations, or knowledge that a phenomenon fits in a unified system of knowledge. Thus, Peter Lipton (2004, p.30) writes: "Understanding is not some sort of super-knowledge, but simply more knowledge: knowledge of causes." If we know, for example, that the global temperature has risen in the past century, we have knowledge of global warming; and if we know, in addition, that the cause of this global warming is the increase of $CO_2$ in the atmosphere, in combination with the greenhouse effect, we have understanding of the phenomenon of global warming. Although this might seem a plausible view at first sight, it can be argued that mere *knowledge* of causes is insufficient for understanding. For it is one thing to *know* that the greenhouse effect causes global warming, but it is quite another to *understand* why or how it does so. A student may be able to answer the question 'Why does global warming happen?' correctly by answering 'Because of the greenhouse effect'. But this does not imply that he understands why global warming happens – he merely knows what the cause is. This suggests that understanding must be more than just having a particular type of knowledge, and that this 'more' is a *skill*: the student understands why global warming happens if he has the ability to apply his knowledge, to see what its consequences are in specific cases. The conception of understanding as skill has been elaborated and defended by one of us (De Regt 2009; 2017), and has gained considerable support in recent years.[8]

We submit that the view that understanding is inherently related to skills or abilities also illuminates the debate about theories without a spacetime, whereas the understanding-as-knowledge view fails to make sense of the controversy. To substantiate this claim, we will analyse the case in terms of the contextual theory of scientific understanding, first presented in De Regt and Dieks (2005) and further developed in De Regt (2009; 2017). The contextual theory starts from the observation that understanding is a pragmatic concept, that is, understanding is a three-term relation between a theory T, a phenomenon P and a subject S (who understands P with T). While for traditional philosophers of science the pragmatic nature of understanding was a reason to reject it as subjective, more recent analyses of

---

[7] See De Regt, Leonelli, and Eigner (2009) and Grimm, Baumberger, and Ammon (2017) for collections of papers by philosophers of science and epistemologists on the nature of (scientific) understanding.

[8] More recent proponents of a skill-based account of understanding are Wilkenfeld (2013), who characterizes understanding as representational ability, and Newman (2012, 2014), who conceives of it as an inferential ability.

scientific methodology entail that understanding is indispensable for achieving the epistemic aims of science. The crucial point is that explanation is not simply an objective relation between an abstract theory and a concrete phenomenon, but always involves the construction of a model that 'mediates' between theory and phenomenon (Morgan and Morrison 1999; Morrison 2015). The abstract theory of quantum mechanics, for example, does not simply supply true propositions about the behaviour of real hydrogen atoms. Essential to theoretical description and explanation of facts about hydrogen is the construction of a suitable model of the hydrogen atom, which requires appropriate idealizations and approximations. It is here that pragmatic understanding plays a crucial role. For constructing the mediating model requires the skills to make the right idealizations and approximations – these skills are a form of tacit knowledge, acquired by scientists during education and in practice; they cannot be replaced by strict rules or algorithms. To capture this requirement, De Regt (2017, p.40) introduces the notion of intelligibility (of a scientific theory), which is defined as:

**Intelligibility:** the value that scientists attribute to the cluster of qualities of a theory (in one or more of its representations) that facilitate the use of the theory.

The central thesis of the contextual theory is the idea that intelligibility (of theories) is a necessary condition for understanding phenomena scientifically, summarized in the Criterion for Understanding Phenomena:

**CUP:** A phenomenon P is understood scientifically if and only if there is an explanation of P that is based on an intelligible theory T and conforms to the basic epistemic values of empirical adequacy and internal consistency.[9]

Note that CUP relates two different types of understanding: the understanding of phenomena and the intelligibility (pragmatic understanding) of theories. While these two types should be clearly distinguished, they are related in the sense that the latter is a necessary condition for the former. As we will show in Section 3, applying this distinction to the debate about theories without a spacetime will clear up much confusion. Since these theories have not yet been experimentally confirmed (apart from some very specific approximations in which they reproduce quantum field theories or general relativity), our main concern in this paper will be with the intelligibility of theories, rather than with CUP.[10]

Of course, the key question is: Which theories are intelligible? First, it should be emphasized that intelligibility is not an intrinsic property of theories, but a contextual value: a scientific theory is intelligible to scientists if their skills are attuned to its qualities. Accordingly, judgments of whether or not a scientific theory is intelligible may change with the historical,

---

[9] De Regt (2017, p.92). This is a strongly revised formulation of the version of CUP presented in De Regt and Dieks (2005, p.150).

[10] This is, of course, not to say that there are no phenomena that could not eventually prove to be explained using theories of quantum gravity. Dark matter and dark energy may well turn out to be such phenomena. However, there is currently no consensus over whether these phenomena are related to quantum gravity, or can be explained by conventional quantum field theories. An alternative approach might be to extend the notion of 'phenomena' to include physical phenomena that are predicted by the theory but have not yet been observed experimentally, or have only been observed in analogous cases. The observation of Hawking radiation in an 'analogue black hole', built as a Bose-Einstein condensate, is still to be replicated by different groups. This kind of analogous experiments have been claimed to provide 'confirmation', in the Bayesian sense, of the Hawking effect for black holes; for a philosophical discussion of the notion of confirmation in analogue models, see Dardashti et al. (2016). But we will not need to endorse these conclusions, since we will limit ourselves to theories.

social or disciplinary context (this is confirmed by various case studies in the history of physics; see De Regt 2017, Chapters 5-7). This does not imply, however, that evaluation of the intelligibility is a purely subjective and arbitrary affair, a matter of taste. The pragmatic and contextual nature of intelligibility is not at odds with the possibility of objectively testing the intelligibility of theories (for specific scientists, in specific contexts). A test that is especially suitable for theories in the physical sciences, is the following Criterion for the Intelligibility of Theories:

**CIT$_1$:** A scientific theory T is intelligible for scientists (in context C) if they can recognize qualitatively characteristic consequences of T without performing exact calculations.[11]

CIT$_1$ can be regarded as an objective test of the skills that are needed to construct models that provide explanatory understanding. Which skills are needed obviously depends on the properties of the theory. These theoretical properties provide the 'tools for understanding' and the scientist should be suitably skilled to use these conceptual tools. By applying CIT$_1$, one can check whether a scientist's skills are attuned to the theory's properties. Because qualitative insight into the consequences of a theory can be gained in many ways, CIT$_1$ can accommodate the variety of ways in which understanding is achieved in scientific practice.

The scientific theories referred to above are not merely formal theories, but interpreted theories. Following De Haro and De Regt (2018: section 1.1), who distinguish between a *bare theory* (a set of laws, or a set of equations) and an *interpreted theory*, our concern here is invariably with interpreted theories, not bare theories. For an interpretation is needed before a theory can make any predictions, represent a physical system, provide explanations or promote understanding. Thus, understanding is one of the aims of interpretation. Also, interpretation is a precondition for understanding: since a physical theory cannot be used fruitfully without an interpretation. Interpretation often proceeds through the development of *tools* that allow scientists to use the theory. As we argued in De Haro and De Regt (2018: section 1), the tools that allow scientists to interpret a theory are typically also tools that render the theory intelligible, in the sense of CIT$_1$. In particular, the tools for understanding that we will list in Section 4.1 are primarily tools for understanding theories without a spacetime: and it is by developing interpretations that the theories in question are rendered intelligible.

A prominent example of a theoretical quality that may enhance intelligibility is *visualizability*. Many scientists employ visual reasoning when constructing explanatory models, using pictorial representations or diagrams as tools. The history and practice of science shows that visualization often serves as a tool for understanding. But success is not guaranteed, as the case of quantum physics shows. Since the advent of quantum mechanics in the 1920s, unambiguous, direct visualization of atomic structure has proven impossible, and subsequent developments have only reinforced the abstract, non-visualizable character of quantum theory. Initially, this raised the question of whether quantum mechanics was an intelligible theory at all. Schrödinger answered in the negative, since he regarded visualizability (*Anschaulichkeit*) as a necessary condition for understanding. His argument was that "we cannot really alter our manner of thinking in space and time, and what we cannot comprehend within it we cannot understand at all" (Schrödinger 1928, p.27). Thus, Schrödinger considered

---

[11] CIT$_1$ is only one of many possible criteria (tests) for intelligibility, which may be termed CIT$_2$, CIT$_3$, CIT$_4$, …

a space-time framework to be a prerequisite for understanding. The contextual theory of scientific understanding explains his preference for direct spatiotemporal visualization by showing how it functions as a conceptual tool that enhances the intelligibility of quantum theory; that is, it facilitates the use of the theory in constructing explanatory models of phenomena. But – contrary to what Schrödinger himself asserted – it does not follow that visualizability is a necessary condition for understanding: alternative conceptual tools may fulfil the same function and a theory can be intelligible without being visualizable.

Let us briefly look at the history of quantum theory, to see how physicists dealt with obstacles to visualization and scientific understanding. Matrix mechanics, developed by Heisenberg in 1925, was the first quantum theory that resisted spatiotemporal visualization completely: it did not offer a spacetime description of atomic structure but merely specified relations between observable quantities (e.g. frequencies and intensities of spectral lines). While physicists from the Copenhagen circle enthusiastically received matrix mechanics, the theory was largely ignored outside this relatively small group. Its lack of visualizability was an important reason, as this hampered application in other domains. Its mathematical intricacy thwarted solution of general cases, and even explaining the hydrogen spectrum turned out to be extremely complicated. In sum, the loss of visualizability engendered a loss of intelligibility that (at least temporarily) reduced the actual problem-solving capacity of quantum theory.

These problems motivated Schrödinger to develop his wave mechanics as an alternative. Representing atomic structure in terms of wave functions, this theory promised spatiotemporal visualization of atomic processes as wave phenomena. Comparing wave mechanics with matrix mechanics, Schrödinger (1928, p.26-27) argued that the former allows for an interpretation that avoids the conclusion that space-time description of atomic structure must be given up, a conclusion that he regarded "as equivalent to a complete surrender". He added that the strength of wave mechanics lies in its "guiding, physical point of view", promising "*anschauliches* understanding" (p.30). Visualizable theories are heuristically more powerful, according to Schrödinger. He saw a relation between visualizability and intelligibility that accords with our analysis: visualizability is a quality that facilitates use of the theory. Indeed, wave mechanics turned out to be more successful than matrix mechanics (Beller 1999: 36-38). Its visualizability and mathematical structure enhanced its intelligibility: it matched the skills of most physicists much better than matrix mechanics. It was easily applicable to concrete situations, and allowed physicists to solve problems that matrix theory was unable to handle. For example, it yielded the spectrum of hydrogen quite straightforwardly. However, direct spatio-temporal visualization of atomic structure in terms of the wave picture turned out to be impossible, because an $n$-particle system is represented by a wave function in $3n$-dimensional configuration space, not in three-dimensional real space.

The years after World War II witnessed a similar development with respect to quantum field theory. In 1948, soon after Schwinger and Tomonaga had developed an abstract mathematical formulation of quantum electrodynamics that resisted visualization, Feynman proposed his famous diagrammatic approach. Feynman diagrams do not represent physical processes but are tools for simplifying calculations. Thus, they enhance the intelligibility of quantum field theory by facilitating its application to concrete problems. The visual nature of the diagrams was a key factor in the success of Feynman's approach, but this success was not immediately achieved. Initially, there was confusion about how they should be interpreted and used (Kaiser 2005, pp. 43-51). Effective use of Feynman diagrams requires specific skills that can only be acquired through apprenticeship and in practice. But some years after their introduction the diagrams were universally accepted and widely used. In sum, Feynman diagrams function as conceptual tools that help rendering quantum field theory intelligible to theoretical physicists. The success of Feynman's diagrammatic method suggests that physicists appreciate visualization as a tool for making theories intelligible.[12] However, as in the case of wave versus matrix mechanics, this does not show that visualizability of the theory itself is a necessary condition for its intelligibility. Understanding may be achieved by other means as well. We conclude that visualization is a 'contingently dominant' tool for understanding: while it is often used and preferred by many scientists, it is not indispensable.[13]

### 2.2. Rejecting metaphysical apriorism

Understanding can also be had without visualization, and without a spacetime in which to visualize. We will, in Section 4, extend that analysis to theories of quantum gravity. But before we can meaningfully do so, we should address an objection, based on 'metaphysical apriorism': the thesis that our knowledge and understanding of the world require a particular metaphysics.[14] Applied to scientific theories, metaphysical apriorism says that, if a theory is not based on a preferred (fundamental) ontology (because it does not contain, or does not allow, the concepts which are associated with that ontology), then it is not a good theory: it is not intelligible, or it cannot give us understanding of the world. In our context, the relevant kind of metaphysical apriorism assumes that scientific understanding of quantum gravity phenomena must be based on an ontology that contains spacetime at the most basic level.

In so far as metaphysical apriorism may display a lack of imagination on the side of its contenders, its appearance in science seems to be straightforwardly explained in Kuhnian terms. Scientists who have been trained within a specific scientific paradigm invariably find it hard to rewire themselves to work within a new paradigm. The limitation is pragmatic, and prompted by the emphasis on a set of paradigm-specific exemplars in scientific education, and by the incommensurability of the different paradigms. Coming to see the world within a new paradigm may entail giving up old metaphysical commitments: and such conversions, from one paradigm to another, are not only a matter of a decision over the best arguments: they involve revisiting ontological commitments which are hard to change. While the point is obvious from the point of view of the history of science, we will give some examples below.

While our knowledge of the world always presupposes some kind of idea about what the world is like, and thus a minimal ontology, it is not the case that our search for knowledge and understanding should limit itself to a once-and-for-all fixed ontology: not even some once-and-for-all fixed aspects of an ontology. Formulating a convincing ontology is one of the main *aims* of the activities of physicists and philosophers, viz. to describe what the world is like, rather than the *starting point* of their activity.

---

[12] See De Regt (2014; 2017, Chapter 7) for more detailed analyses of these historical cases. Kaiser (2005) presents an extensive account of the history of Feynman diagrams and analyses their function as 'paper tools'.
[13] Cf. Cushing (1994, p.20-21) on the principles that govern human thinking: "they have the status of contingent but universal principles of the human mind". For further philosophical discussion of the indispensability of visualization for scientific understanding, see Möβner (2015).
[14] Knowledge and understanding always assume *some* metaphysics; so, by 'particular', we here mean that this metaphysics has to be of a certain kind, and that it is specified beforehand: so that a metaphysics of a different kind cannot provide knowledge or understanding.

How scientifically unproductive it is to do science with an ontology that is not open to further modification, is illustrated by the failure of the Cartesian programme of physics based on the mechanistic principles of the extension and impenetrability of corpuscles, and of their interaction by contact. Had Newton strictly followed Descartes' mechanistic philosophy, he would not have even possessed the conceptual tools needed to formulate his law of gravity, which acted at a distance and not by contact (see De Regt 2017, Chapter 5).

In view of these criticisms, theoretical and pragmatic, one may retreat to the position that metaphysical apriorism is not based on an assumption about what the world is like, and so it does not attempt to impose an ontological condition on scientific theories, but about the structure of our mind. On this view, the human mind is wired to work within the limits of certain concepts (in particular, the concept of Euclidean space), whether the world in itself fits the description in the mind or not. But the tenability of such views has been eroded by the successful application of non-Euclidean geometries in physics. There is no doubt that humans can understand theoretically, and can work practically with, non-Euclidean geometries, thus gaining understanding of phenomena from them.

In Section 3, we will discuss authors who stress the importance of certain types of spacetime concepts for understanding: Maudlin (2002), and Lam and Esfeld (2013). To be sure, these authors do not manifest themselves as metaphysical apriorists, as they do not assume a preferred metaphysical position from which science has to be done. Instead, they make claims such as: 'it is hard to see how to do physics without spacetime', and 'theories without a spacetime are crazy'. These statements refer to intuitions or pragmatic necessities, rather than to metaphysical necessities. Although Lam and Esfeld do wish to secure an 'ontologically serious' position, their argument for it is pragmatic—it is *hard to see* how a theory without spacetime can be combined with their interpretation of quantum entanglement. But what is the force of such arguments, once metaphysical apriorism has been rejected? First of all: reasonings like "it is hard to see how…", or "how else could it be …" are, in general, suspect philosophical arguments, and make for unconvincing philosophical positions: for they are easily attributed to either intuitions instilled by training or to a lack of imagination. Such arguments are normally superseded by history (as we have shown in the examples above).

More substantially, once we have rejected metaphysical apriorism, what grounds are left for arguing that spacetime should be an element in our theories that is indispensable for scientific understanding? Indeed, it seems that no grounds for that thesis are left at all. It is just as plausible (certainly, if our best physical theories would so compel us to conclude) that the concept of spacetime is simply an approximation to something more fundamental, of which our theories can give us understanding. Thus, metaphysical apriorism may not in fact be yet dead, but it is certainly a dead end.

## 3. The 'Spacetime Wars': Debates over scientific understanding?

In this section, we will address the third question following from (BQ): (iii) Can recent claims and debates in the literature on the philosophy of spacetime be rephrased as definite answers to, or positions on, (BQ)? We will look at two discussions in particular: first, the debate between John Earman and Tim Maudlin over 'frozen' spacetime, viz. a spacetime in which no physical quantities change over time (§3.1); second, recent discussions of emergence of spacetime, in which spacetime does not figure as a concept in some central fundamental physical theories (§3.2). In §3.3 we show that the underlying issue in these debates is our basic question (BQ): Can theories without a spacetime be intelligible? Showing that intelligibility —besides other questions such as explanation—is at stake in these debates will further motivate our analysis of understanding in quantum gravity in Section 4.

## 3.1. Understanding or absurdity? Earman vs. Maudlin

In a paper entitled "Thoroughly Modern McTaggart", John Earman (2002) revisits the status of time, and of change, in general relativity, formulated in the Hamiltonian formalism, i.e. Hamiltonian general relativity, which we will for short write as HGR. It is, of course, not our aim to assess the cogency of Earman's analysis here, or to take a position on the details concerning the physics of this debate (in section 3.1.2 we will mention the reception that the debate has had in the recent literature). Rather, our interest lies in analysing the extent to which the authors deem the absence of a (conventional) view on time to be problematic, from the point of view of scientific understanding.

Earman's conclusions can be summarised in three theses:

(1) There is no physical change in HGR because the theory "implies that no genuine physical magnitude takes on different values at different times" (p. 2)[15]: there are no (or very few) proper observables in general relativity.
(2) However, other, non-standard observables can be constructed which satisfy the criteria of general relativity. Earman calls these observables 'point coincidences' (also often called 'relational observables').
(3) Because of the existence of relational observables, HGR "does not imply a flat-out no change view: it is compatible with an ontology consisting of a time ordered series of occurrences or events, with different occurrences or events occupying different positions in the series." Because of the existence of such a time series, the view does not entail that time is not real. The idea is, rather, that temporal change "is not to be found in the world in itself but only in a representation" (p.14).

Physical change (1) would occur if some genuine physical magnitude or observable were to take on different values at different times.[16] But it follows from Earman's analysis of the notion of observable in HGR that no (or very few) such observables exist.[17]

But Earman also has a soothing for the troubled reader: there is no reason to despair, for one can construct other quantities, called 'point coincidences' (2), which will serve as observables; and in terms of *these*, a time series (3) can be constructed. This new series is not constructed out of the *primary* quantities of general relativity (1), but out of the *derived* quantities (2). Thus change only exists in a 'representational' sense (3), i.e. not as changes of the primary objects but only as changes of their representations. Thus, change is no longer a property of the world 'in itself': it is only an appearance, but one based on an objective structure of the physical world, and on our best physical theories about it.

---

[15] Earman assumes McTaggart's B-series account of time, on which events are ordered by an earlier-than relation. According to the alternative A-series, events are ordered as past, present and future. Further details of the B and A series (introduced by McTaggart in 1908) will not be important for our analysis.

[16] The claims, in the physics literature, that these theories lead to the absence of observables that change in time, or lack of a fundamental spacetime, are of course older. For instance, Earman (2002) makes key use of Bergmann's and Komar's work from the 1950s and 60s.

[17] The argument is as follows: Earman (following Bergmann) proposes to resolve the indeterminism implied by the hole argument by demanding that what we call 'observables' be diffeomorphically invariant quantities. But then none of these quantities can change in time. For instance, if one considers the scalar curvature as an observable, it then follows from diffeomorphism invariance that the scalar curvature must be a constant over the entire manifold. A similar argument applies to more general 'quasi-local field quantities', which are obtained by integrating a local field quantity over a spacetime region. See Earman (2002: p. 10).

Before moving on, we should briefly discuss in what sense Earman defends a theory 'without a spacetime'. Compared to the views which we will discuss in Section 4, it is only a weak version of such a view. From Earman's analysis (1)-(2), it follows that the time variable of general relativity in the Hamiltonian formalism is only part of what he calls the 'surface' structure of the theory, but not of its 'deep structure' (p. 3). As we have seen, there is only an *appearance* of change. So, as interpreted by Earman, general relativity is a theory without what we might call 'fundamental time'. And this is what Maudlin takes issue with.

The title of Maudlin's (2002) reply, "Thoroughly Muddled McTaggart", is telling of his aim: "John Earman has conjured up yet another monster to trouble poor old Einstein… I hope to drive a stake through the heart of the undead McTaggart and end his new rampage before it has begun" (p. 1). After praising Earman for his ability at making mathematical physics relevant to philosophers of physics, he goes on to explain the reasons for his criticism: "the motor of this project is supposed to be a very, very surprising feature of the "deep structure" of GR: namely that according to the deep structure, nothing physically real changes. This, and only this, is the claim I seek to demolish. Even if I succeed, much of interest may be found in the *disjecta membra* of Professor Earman's paper" (p. 1).

Here is Maudlin's central fear of Earman's monstrous creation: despite Earman's careful interpretive remark (3), that his is not a "flat-out no change view" and that there *is* a notion of change, there is, according to Maudlin, *no real physical change.* And the lack of physical change is a consequence of the avoidance of the indeterminism which Earman has introduced into the formalism of HGR. Maudlin reviews this "unphysical indeterminism" of the dynamics, which is a consequence of the way gauge freedom is taken into account by HGR, and Earman's treatment of it.[18] And Maudlin's verdict of Earman's technical methods leaves no room for doubt: "All of this is, of course, both *formally correct* and completely *crazy*… if we blindly demand determinism from quotienting [gauge freedom], it can certainly meet our demand, but perhaps in a rather *nonsensical* way" (p.7: our emphasis). The emphasised words carry what, from our interest in this paper, stands out the most in Maudlin's criticism: he bemoans the fact that, though the theory is formally correct, it is nevertheless *completely crazy and, indeed, nonsensical.* The rhetoric is strong: in another place, he remarks that "this indeterminism is *completely phony*: it has nothing to do with any real physical indeterminism" (p.9).

In his reply to Maudlin (2002) in the same reference, Earman makes the point of disagreement yet more vivid: "Tim does a brilliant job of explaining the guts of some difficult technical issues. He takes his explanation to show that the sorts of considerations I adduced in favour of modern McTaggartism lead to *a precipice below which lies absurdity. I see no precipice but*

---

[18] In the Hamiltonian formalism, gauge transformations are identified as the transformations generated by the first-class constraints (Earman (2002: p. 8)). However, we will not need to consider these technicalities for our aim.

*rather a series of steps that lead to an understanding* of the motivation and content of contemporary main-line research in the foundations of classical general relativity theory and quantum gravity" (p.19: our emphasis).

We will argue below that we have here a clear case of a debate, within the context of sophisticated theories in mathematical physics, of differing ideas over how the theory is supposed to lead to understanding, with the consequent implications for what counts as good guiding principles for new research. Maudlin does not question the formal correctness of Earman's mathematical procedures, as the above quote shows (p. 7); rather, he emphasises the role of having a theory in which there is "real physical change in the world" (p. 13), not a theory that needs to be interpreted "through the looking glass", and in which "we are apparently to accept some Alice-in-Wonderland logic" (p.13). Indeed, Maudlin would like to measure time in simple and concrete ways: through "the precession of the perihelion of Mercury or the reading of our clocks" (p.15), or through the simple coincidence of geodesics (p.18): "by sending a rocket along each path and making the measurement when they collide" (p. 18). But Earman's sophisticated analysis (1)-(2) would seem to forbid such simple procedures from counting as 'observables', in the technical sense.

Earman characterises Maudlin's stance as appealing to intuitions: "Suppose that the result offends your (or Tim's) intuitions… But in the absence of any competing method… you (or Tim) should seriously consider the possibility that your intuitions have to be retrained." (p.19). For Earman, a priori physical intuitions about determinism or measurement in HGR do not have a special status when confronted with new theories: "it is hard to see any principled way to distinguish tolerable vs. intolerable violations [of determinism] of this kind" (p.20). And so, the disagreement between Earman and Maudlin is about what counts as absurd and what does not: "all of the leading research workers in this program… accept… the consequence that the observables of GR are 'constants of the motion', a consequence that Tim labels as absurd and disastrous" (p.20-21). The final paragraph of Earman's paper again contrasts what Maudlin regards as an 'absurdity' with what, according to Earman, is only 'understanding': "If I could make Tim feel the excitement I experience when I see how philosophical concerns about time and change intertwine with contemporary research in physics, he might be willing to join me on the precipice—a precipice not of absurdity but of a new understanding of old issues." (p.23).

The reception of Earman's arguments in the philosophy of physics has been mixed: some commentators tend to side with Earman (e.g. Huggett, Vistarini, and Wüthrich 2012: p.7)),[19] while others are critical about Earman's use of general relativity in the Hamiltonian formulation (e.g. Pitts 2014; Gryb and Thébault 2014). Pitts (2014) argues that upholding the equivalence of the Hamiltonian and Lagrangian formalisms resolves the paradox. He suggests that time, and thus change, can be upheld in general relativity. We will here not engage in an evaluation of these arguments (nor do we need to, as we will now argue). Notice that Pitts' expressed sympathy for Maudlin's critique does not mean that he takes Maudlin's arguments to be convincing ("beating back Poisson brackets with appeals to common sense does not yield full conviction, and rightly so", p. 3).

---

[19] These authors discuss both positions in some detail, but their cards seem to lie with Earman, whose case they call "compelling" (p. 7): "one cannot easily escape the grip of Earman's argument" [to the extent to which one takes HGR seriously, which these authors do] (p. 8). Their motivation is that, although they recognise that the Hamiltonian and the standard formulations of GR are not equivalent, HGR may be vindicated by its services to quantum gravity (p. 8). And so, they conclude that Maudlin's proposed solutions do not work. "Maudlin laments that this will lead to the "rather silly"—indeed "crazy" and metaphysically monstrous—conclusion that there cannot be change… Such lamentation, however, does not amount to a counterargument—it is a restatement of the problem! Earman's modus ponens is Maudlin's modus tollens" (p. 8).

An analysis of the debate as a problem of mathematical physics would attempt to settle the question of whether HGR and standard GR are equivalent. But our concern is not about the physics,[20] but rather about the reasons for which Earman thinks that HGR is perfectly intelligible, while Maudlin says that (at least on Earman's treatment) it is "absurd" and "crazy", hence unintelligible. Our main issue is Earman and Maudlin's differing demands on the interpretation of HGR, which we will discuss in the next section.

One might be inclined to argue that the debate between Earman and Maudlin is a debate about the empirical import of the theories: for example, about the empirical coherence of HGR with the way in which we measure time. And we agree: part of the discussion is about this. But we also submit that, behind these surface disagreements about physics and empirical matters, there is a deeper disagreement about the intelligibility of the theories. If the disagreement were a 'mere' fact of empirical coherence, it would be hard to understand the rhetoric of the debate. For the authors do not hesitate to use terms like 'absurdity', 'nonsense', 'crazy', 'phony', and 'guts'. Clearly, the stakes are high: and it seems that a charitable reading of such strong rhetoric must not assume that the disagreement is merely about cognitive matters that can easily be settled in a scientific discussion (which would hardly justify the repeatedly used strong language). Rather, it seems that we must view the debate in Kuhnian terms, i.e. in terms of the use and pragmatic aspects of the theories—and this prompts us to reframe the debate in terms of scientific understanding. We will now cite additional evidence for this reframing.

Our first cue is Earman's *own* use of the word 'understanding' at central places in his brief reply to Maudlin: "I see no precipice but rather a series of steps that lead to an understanding of the motivation *and content* of contemporary main-line research in the foundations of classical general relativity theory (GTR) and quantum gravity." (p. 19: our emphasis). Understanding here refers to both the motivation *and the content* of research in general relativity: hence, in Earman's own words, the *intelligibility of general relativity,* in its Hamiltonian formalism, is surely at stake here. Also, in the final sentence on p. 23, Earman uses the word 'understanding' in a similar sense, namely as "understanding of old issues" in general relativity. And this use of 'understanding' is underlined by the vivid contrast, earlier in that same sentence, with the "precipice of absurdity" that Maudlin thinks Earman's work leads to: which further supports our literal reading of 'understanding'.

Further supporting our interpretation is provided by Earman's suggestion that his claims have offended Maudlin's "intuitions", and that intuitions "have to be retrained". This suggests that his use of 'understanding' must be taken in a pragmatic sense; it concerns understanding of theories and standards of understanding, and not questions that can simply be settled in a scientific debate. This is why the debate is more than a disagreement about physics, or even about the interpretation of general relativity:[21] it is also a debate about how scientists ought to *learn to work with* HGR.

---

[20] Even if HGR turned out to be inequivalent to standard GR, there might be other motivations for attempting to interpret this theory. For example, Huggett, Vistarini, and Wüthrich (2012) find motivation in its use for quantum gravity.

As we already mentioned, words like 'absurd', 'crazy', and 'Alice-in-wonderland logic' have a subjective flavour, which again more straightforwardly points to an underlying disagreement about understanding than to, say, a mere disagreement over the physical explanation, ontology, or even empirical coherence. Thus, to understand the debate and its rhetoric, we must appeal to this deeper level of what is deemed reasonable and acceptable for scientific understanding. This debate is an interesting rehearsal of its quantum version, where the broad consensus is that the problem of time (as well as the problem of space!) is serious, and where the positions and the arguments have indeed been rehearsed, as we will now show for the case of quantum gravity.

**3.2. Emergence of spacetime in quantum gravity**

In quantum gravity, it is often claimed that space and time are somehow derived, emergent quantities, rather than basic concepts. Huggett and Wüthrich (2013), in a special issue of *Studies in History and Philosophy of Modern Physics* devoted to "The emergence of spacetime in quantum theories of gravity", explain it thus: "While different approaches to quantum gravity are often based on rather different physical principles, many of them share an important suggestion: that in some way spacetime as we find it in our existing theories is *not* a fundamental ingredient of the world, but instead, like rainbows, plants or people, 'emerges' from some deeper, non-spatiotemporal physics… The idea that the universe and its material content might not, at bottom, be 'in' space and time, that these seemingly fundamental ingredients are just appearances of something more fundamental, would, if borne out, shatter our conception of the universe as profoundly as any scientific revolution before." (p. 273). As Huggett and Wüthrich stress, many of the current approaches to quantum gravity share the suggestion that space and time are not among the fundamental ingredients of the world.

But not all seems to be well with this idea. For, in the same volume, Lam and Esfeld (2013) critically assess the fundamental assumption of the special issue: "This paper aims to investigate this radical suggestion [that spacetime may not be fundamental] from an *ontologically serious* point of view… For any known ontologically serious understanding of quantum entanglement, the *commitment to spacetime seems indispensable*… It is *unclear how to make sense* of concrete physical entities that are not in spacetime." (p.286: our emphasis).

Lam and Esfeld defend their claim, that quantum entanglement carries an indispensable commitment to spacetime, with two main arguments. First, they argue that an analysis of entanglement in non-relativistic quantum mechanics implies the indispensability of spacetime, i.e. that it presupposes spacetime: "The quantum systems are separated in space; against this background, they are non-separable in the sense that they do not possess separate states each. Consequently, this ontology presupposes that the systems to which it applies are inserted in spacetime and are somehow localized in spacetime." (p. 289). Their view gains further support from their analysis of the non-local dynamics for these systems. "There is no sense in which there could be concrete physical structures of entanglement… unless they are implemented or instantiated in spacetime." (p. 291). By itself, this argument does not rule out the fundamental absence of spacetime: "These considerations do of course not constitute an argument against QG [quantum gravity] being committed to entities that are ontologically more fundamental than spacetime. They only show that if there are such entities, it does not make sense to apply notions such as entanglement or non-separability to them." (p.291).

---

[21] If 'interpretation' were to be used in a narrow sense devoid from pragmatic issues, that is. But, as we mentioned in §2.1, interpretation is a precondition for understanding. Thus a problem of interpretation gives *ipso facto* also a problem of understanding. In fact, the connection is here more direct, because Earman's appeal in 'retraining intuitions' is an appeal to *pragmatic* aspects of interpretation, which are directly relevant for understanding.

Second, from Lam and Esfeld's 'ontologically serious point of view', it is hard to see how physical entities could fail to be in spacetime. Their ontological standpoint is that of *ontic* structural realism (OSR): "OSR has to be precise about what it takes the concrete structures to be… This requirement implies the commitment to spacetime: the structures that OSR admits are concrete physical structures through their being embedded, implemented or instantiated in spacetime. Without the commitment to spacetime, it would simply be unknown as in ESR [epistemic structural realism] what the entities are that implement or instantiate the mathematical structure of the theory in question." (p. 289).

Their arguments often employ premises like "it is unclear that…" and other similar phrases, which appear a total of 11 times in their paper. Related expressions like "it is difficult to see how…" appear another five times. Invariably, the meaning of these phrases is that the authors cannot make sense, within their "ontologically serious point of view" and their views about entanglement of quantum systems and physical causation, how emergence can be applied to entities which are not spatio-temporal. On the other hand, quantum entanglement, with its reference to spacetime, "provides for a clear and serious ontological view" (p. 289). Like Maudlin, Lam and Esfeld appeal to the necessity of having concrete, physical, understanding, according to spacetime categories drawn from previous theories that are deemed indispensable. For these authors, spacetime crucially enters the distinction between physical and mathematical structures: "quantum structures need to be implemented in spacetime in order to be concrete physical structures in contrast to abstract mathematical ones" (p. 292).

### 3.3. The problem of spacetime as a problem of understanding

As we have seen above, the notion of 'understanding' plays a central role in the debates about theories without a spacetime. Apparently, the stakes are high, as the opponents of such theories do not hesitate to use terms like 'absurdity', 'nonsense', and even 'crazy'. Maudlin's charge is that they are unintelligible and cannot provide any understanding, while Lam and Esfeld claim that theories without a spacetime are at odds with what they call an "ontologically serious point of view". In order to make sense of these debates, and to determine which of the arguments are merely rhetoric and which do actually carry scientific or philosophical weight, we will analyse the arguments from the perspective of recent philosophical debates about the nature and function of understanding in science. The discussion above should have made clear that the necessity, the convenience, or the possibility, of a description of nature without spacetime are defended or attacked from different stances on how one is to understand such theories, and how such theories lead to understanding. More precisely, the positions correlate with the two possible answers to question (BQ), which we will call the 'formalist' and the 'intuitionist' positions.[22]

---

[22] Our use of the labels 'formalist' and 'intuitionist' does not rely on any previous use of these words. Rather, our use of these labels refers to the factors that are given preference by the two positions: a well thought-out formalism of a theory, in the case of the formalists; or some pre-existing intuitions about the world (about

'Formalists', such as Earman, as well as all of the physicists and philosophers of quantum gravity who defend the approaches which lack a fundamental spacetime, are happy to follow their mathematically formulated theories to whatever conclusions these will lead to, as long as these conclusions are carefully thought-out and lead to physical interpretations—but these interpretations are not to be held accountable to strong pre-conceived ideas about spacetime.[23] As Earman eloquently puts it, he is happy even in the face of what appears to some to be "a precipice below which lies absurdity". For, to him, there is "no precipice but rather a series of steps that lead to understanding". Earman is happy to retrain his physical intuitions if his most successful theories require him to do so.

'Intuitionists' such as Maudlin, Lam and Esfeld, counter that such theories are not intelligible, and reject them on *those* grounds. There are intuitive considerations of various kinds which must be upheld in the interpretation of new theories—which can thus also function to reject theories, or interpretations of theories, which do not accommodate to the standards set by those intuitive considerations (that there are such standards is indicated by, e.g., the use of the word 'serious' by Lam and Esfeld (2013); and in Maudlin's strong language of 'phony', 'Alice-in-wonderland logic' etc.). Such intuitive considerations can come from various sources: in the case of Maudlin, the everyday and the scientific observations of the reality of time and of change provide such a basis.[24] For Lam and Esfeld, metaphysical considerations, as well as the achievements of previously established scientific results (like entanglement of quantum systems) provide such grounds.

It is important to notice here that the intuitionists do not present a knock-down argument, i.e. an argument from the mathematical inconsistency or the empirical inadequacy of the theories (recall Pitts' (2014: p. 3) remark that Maudlin's common sense is not enough to beat back Poisson brackets; Maudlin himself explicitly says that Earman's treatment of the theory is "both formally correct and completely crazy", p. 7). Rather, matters of *interpretation and physical expectation* are at stake. Furthermore, the intuitionist argument relies on judgment about what the reliable sources are for interpreting theories (common sense observation, metaphysics, etc.). The intuitionist arguments thus appeal to further physical intuitions. Clearly, the intuitionists hold on to a logically stronger, because more restricted, standard of scientific understanding. They do not consider Earman's admonition to "retrain intuitions" (2002: p.19) to be attractive or convincing: for this would imply giving up their pre-theoretical conditions—in particular, giving up the understanding of nature in space and in time. In their view, this would imply a weakening of the standards employed in prescribing how a theory should explain the world.

It thus follows that the intuitionists deem the theories without a spacetime either incapable of giving understanding of certain key phenomena; or they deem the theories without a

---

change in the world, about metaphysics, etc.), which take preference over specific theories, in the case of the intuitionists.

[23] For more on the notion of interpretation, see De Haro and De Regt (2018: section 2.1).

[24] An anonymous reviewer suggested that Maudlin's position is not based on mere intuition but rather on the argument that the observable phenomena to be explained are always spacetime phenomena. This argument refers to what Huggett and Wüthrich call empirical incoherence, and as we mentioned in Section 3.1 this indeed plays a part in the debate. However, the argument also concerns the lack of a connection with our experience and with the working of our measuring instruments. According to Maudlin (2002, p.12), Earman's interpretation "separates our experience from physical reality". Here Maudlin's 'intuitive' considerations regarding the intelligibility of theories come to the fore.

spacetime themselves to be unintelligible. It is important to notice that the appeal here is to *understanding* rather than to *explanation*. For the disagreements that we have highlighted are not only about any causal or mechanistic accounts of phenomena, or just about concrete puzzles raised by these theories, but rather about the lack of certain properties which render the theories intelligible, or as not leading to understanding—and this is what explains the strong language used.

## 4. How theories without a spacetime can be made intelligible

In this section, we will first list some tools that are available for interpreting theories, and thus for rendering them intelligible (§4.1). Then we will discuss, in §4.2, examples of models in which an *effective spacetime* is developed, even in cases in which a fundamental spacetime is absent, thus gaining understanding through a particular type of visualization. Effective spacetimes act as crutches that physicists use, and on which they build understanding, even if effective spacetimes are not always interpreted as the spacetime we live in, but simply as auxiliary structures. In §4.3, we will present examples of models in which physicists have not, in fact, developed (or have not needed to develop) a spacetime of any kind, yet have gained understanding through the use of the toolbox in §4.1.

### 4.1. Cataloguing the tools

In this section, we will discuss which tools can be used for understanding in theories that do not involve a fundamental spacetime. We will analyse the kinds of tools that may be expected to render such theories intelligible. Some of these tools will be geometric and thus involve visualization, others will not. In the companion paper De Haro and De Regt (2018) we elaborate these alternative tools for rendering theories intelligible in more detail.

Broadly speaking, we can distinguish three kinds of tools:

(Similar) Similarities between the functional role that theoretical concepts play in different theories or models. For instance, Hilbert spaces and gauge groups appear both in quantum mechanics and in quantum field theory. Depending on the examples, these similarities range from actual identities (e.g. the same Hilbert space appearing in two theories and playing exactly the same role) to analogies: where the differences lie either: (i) in the mathematical object itself, e.g. a different gauge group is interpreted in the same way; (ii) in its interpretation, e.g. different functional roles for the same Hilbert space.

(Internal) Internal criteria, based on the role that the concepts play within the theory: indeed, once some elements of the theory have been given an interpretation which allows physicists to grasp those elements, now other elements of the theory can also be interpreted, based on their functional connection to those other elements within the theory. In this way, the theory as a whole becomes more intelligible. Also, a given mathematical structure, as a whole, can receive an interpretation using other methods: and then the concepts in the structure receive their interpretation from their overall role within the structure, thus rendering the theory more intelligible.

(Approximation) Use of approximations: new theories (also theories without a spacetime) are supposed to reproduce known phenomena and theories in specific approximations, often given by limits. These are interpretations 'from below': interpreting the lowest elements within the theoretical structure (say, an effective spacetime as compared to our physical spacetime), the theory can receive an interpretation at other scales as well. Thus approximations, in so far as connected to effective spacetimes, usually involve visualization.

In the next section we will give examples of the use of the above three criteria. It is important to note here that these criteria usually work together, rather than in isolation. For instance, once a theory has been interpreted in a particular approximation, then some concepts in the higher-level theory receive an interpretation, but not all: and so, one needs a combination of (Approximation) and (Internal) to interpret those remaining elements. Understanding a theory normally proceeds by iteration of these three tools.

**4.2. Visualization via effective spacetimes**

As we have argued in §2.1, visualizability is a contingently dominant tool for understanding. The dominance of this tool is illustrated by the fact that, even in theories which do not postulate the existence of a spacetime at the fundamental level, spacelike and timelike structures do appear at some level of analysis. We will call these structures, resembling spacetime, 'effective spacetimes'. This often happens in the case of (Approximation), as we will illustrate in this subsection. The appearance of effective spacetimes in the theory can have two kinds of interpretations, which help rendering the theory intelligible: (a) Physical: effective spacetimes reproduce the spacetime we live in, or something similar to it, in a suitable approximation. (b) As auxiliary mathematical constructions which do not represent a literally physical spacetime, but aid our physical intuition, which can more easily reason using geometric notions. The physical nature of the former kind of effective spacetime is often associated to its emergence from a theory in which there is no spacetime.[25] Here, this distinction will not be relevant, because both (a) and (b) contribute to intelligibility.

Our examples of emergent spacetimes in this subsection will be of two kinds: (§4.2.1) effective spacetimes in theories which do start off with a spacetime (though the latter is often not interpreted as 'physical'). In this case, the effective spacetime can either live alongside the original spacetime, or *replace it*. These examples are more familiar and numerous in the physics literature, and so they will illustrate our point well. §4.2.2 will deal with effective spacetimes arising in theories which start with *no* spacetime at all. We take up these two kinds of examples of visualization in turn (for more details, see De Haro and De Regt 2018).

**4.2.1. Effective spacetimes in theories with spacetime**

An example of an effective theory which begins with a spacetime, and in which a *new* spacetime is developed (which comes to replace the old spacetime), is 't Hooft's 'large $N$' analysis of Yang-Mills theory (this is the theory of the strong interactions). In the particular limit of the theory which 't Hooft (1974) considers, strings emerge, and with them emerges an associated spacetime in which the strings move. 't Hooft considered the theory of the strong interactions with a gauge group U($N$), where $N$ is the number of quark colours (equal to 3 in the standard model, but now generalised to an arbitrary number of quark kinds). 't Hooft discovered that, in this particular limit, the Feynman diagram expansion of certain models can be rewritten as a topological expansion of two-dimensional surfaces (rather than the one-dimensional skeletons that are the Feynman diagrams). In fact, the amplitudes of the theory in this limit have the same structure as those of a string theory, in which the Feynman diagrams have now been transmuted to string world-sheets moving in an ambient space which could potentially be curved, and differ from the original flat Minkowski space in the basic ontology of the theory. So, the 'ordinary' spacetime of the theory has disappeared, and has given rise to another, emergent spacetime, whose properties are different from the original one. For a philosophical discussion of this kind of emergence, see Butterfield and Bouatta (2015).

---

[25] Emergence is here understood as novelty of the spacetime, compared to its comparison class (see Butterfield (2011), De Haro (2015)).

't Hooft's idea of the relation between gauge theories and strings has been generalised in so-called gauge/gravity dualities, currently a thriving area of quantum gravity research. These dualities are very general relations of equivalence between such quantum field theories on fixed spacetimes, and theories of gravity/string theories on curved spaces (for a discussion of this kind of theoretical and physical equivalence, see De Haro (2016) and De Haro, Teh, and Butterfield (2016)). Other examples of concrete phenomena in models of this kind go under the name of 'geometric transitions': one starts with a theory with a certain kind of spacetime, and in a suitable limit of the parameters the old spacetime disappears, and a new spacetime emerges, with physical quantities being described by a new theory.

The 't Hooft limit of gauge theories involves the tool (Similar) in §4.1. Indeed, the gauge group $U(N)$ is the starting point of 't Hooft's large $N$ duality. Since physicists already have some intuitions about the role of these structures ($SU(3)$ e.g. being the group of the 'colour charge' of quarks), they can now grasp the similar role which these groups play in the new models which they consider (as e.g. a generalisation of the strong interactions). And the appearance of the stringlike structures themselves is, of course, (Similar) to other models of strings with which physicists are already familiar. Both are cases of analogy. But the role of the limit itself, and of the appearance of spacetime in this limit, is clearly a case of (Approximation). In this approximation, we have a familiar theory (a theory of strings), and that interpretation helps physicists organise the information and work with the theory further, thus gaining insight into the original theory.

### 4.2.2. Effective spacetimes in theories without a spacetime

In this subsection, we give the second set of examples, of the genuine kind in which one starts with a theory which has no spacetime in its basic ontology, and develops an effective spacetime.

There is a very wide class of so-called 'random matrix models', in which the fundamental variables of the theory are not fields defined on a spacetime background (as is usually the case in quantum field theory, see §3.2), but are rather *matrix-valued*, i.e. they are sets of real or complex numbers (the classic reference is Brézin et al. (1978)), with no underlying space or time. The fundamental fields of the theory are thus matrices of specified rank (and with specified properties, such as e.g. hermiticity, or unitarity). In some of the models, these matrices are time-dependent, and so time is part of the basic ontology in *those* models. But it is a common feature to *all* matrix models, that there is no space in the theory thus formulated. In many other matrix models, there is indeed neither space *nor* time (see e.g. Brézin et al. (1978: §3)). Space, or spacetime, appears only after a phenomenon that is analogous to a phase transition in statistical mechanics: in the limit in which the rank of the matrix goes to infinity, the theory displays geometrical structures which were not there before. In that limit, the fields of the resulting theory are no longer ordinary matrices, but rather matrix-valued fields defined in spacetime: a quantum field theory description emerges (for a philosophical treatment of this example of emergence of space, see De Haro (2017a)). Spacetime has appeared in a theory in which there was no spacetime to begin with. In one of the versions of matrix models (this is the celebrated 'M theory conjecture', where M stands for 'matrix': see Banks et al. (1997)), the structure of the matrices (which are, in this case, time-dependent) is so rich that an entire theory of supergravity in 11 dimensions, with a quantum version of it—the so-called M theory—arises. For a simpler version, in which the matrices are time-independent, see e.g. Dijkgraaf and Vafa (2002), and Mariño (2004) for a review. Rather surprisingly, the simple dynamics of the matrix eigenvalues transmutes to give explicit examples of so-called 'special geometry', namely a deformed six-dimensional Calabi-Yau geometry, on which strings interact.

The effective spacetimes which arise in these theories are used by physicists as a tool for understanding the theory, as in (Approximation) of §4.1. Indeed, recognising that a well-defined geometrical structure emerges, in an appropriate approximation or limit, allows physicists to recognise a new formulation in which the theory is best framed, often in terms of a familiar theory, such as a string theory. One can then use the new formulation of the reformulated theory to carry out calculations which would otherwise be hard to do in the original theory, but are now easily organised, if one thinks of them in geometric terms. These reformulations are called 'dualities' (for a philosophical account of dualities, see De Haro (2016), De Haro and Butterfield (2017)). Dualities thus have a pragmatic aspect, in that they allow physicists to work with theories that would otherwise be rather intractable, hence increasing the intelligibility of these theories. Indeed, the presence of dualities seems to be one of the factors which physicists think make a theory intelligible, because of its different descriptions, which allows for a variety of views on the same systems. It is thus not surprising that dualities have received so much attention in the physics literature in the last few years—and, by itself, this gives further confirmation to our thesis that dualities render theories intelligible.

By themselves, the examples of intelligibility discussed so far suffice to block the claims, by Lam and Esfeld, that theories without a spacetime formulation cannot be properly understood; or perhaps, we should rather say that they dissolve Lam and Esfeld's perplexity about how understanding can be achieved in theories without a fundamental spacetime. The way in which the above examples dissolve perplexity should be obvious: for understanding of theories without a spacetime is achieved by deriving an effective spacetime, as in (Approximation), which is not fundamental but can be used for physical reasoning. Even if the basic ontology contains no spacetime, geometric intuition can still be used at the effective level, thus contributing to understanding of the theory. Furthermore, structural similarities like (Similar) are also available as tools for understanding.

But the role of spacetime in the examples (b) might not be sufficiently 'ontologically serious', in the sense in which Lam and Esfeld would wish it to be. For the effective spacetime is not part of the fundamental ontology of the theory, but rather a derived entity, which figures at some higher level of the description. And, as we argued in §2.2, it need not be, for metaphysical apriorism is misconceived. As we have argued, understanding of theories which have no spacetime in their fundamental description is possible: through effective spacetimes which appear at other levels.

**4.3. Physics in the abyss: no spacetime and no visualization**

In the previous subsection, we gave several examples of theories which have no spacetime at the basic level, but in which some kind of spacetime is recovered in a suitable limit, thus contributing to understanding through the use of geometric reasoning and intuition, according to the tools (Approximation) and (Similar) in §4.1. (Again, the examples here are further developed in De Haro and De Regt (2018)).

Though we have by now diffused the basic perplexity of Maudlin, and of Lam and Esfeld, by explaining how such theories can be rendered intelligible, our argument would be incomplete if all understanding ultimately always made use of the existence of a spacetime, whether fundamental or effective. Thus the question now becomes pressing, whether understanding can be attained in other ways than by recovering effective spacetimes. In this subsection, we will show that the answer to this question is affirmative, by displaying explicit examples which use (Similar) and (Internal), cf. §4.1, for attaining understanding, but in which an effective spacetime is not present or not needed.

The basic case consists of theories in which, at the basic level, there is no spacetime but something more fundamental: often, a discrete structure. In the causal set approach to quantum gravity, for example, spacetime is fundamentally discrete, and spatiotemporal relations are replaced by a more fundamental partial ordering of events (for a philosophical account, including an elementary introduction, see Wüthrich and Callender (2016)). Here, rather than speaking of an (Approximation) (which is indeed not needed), understanding is already gained directly through (Similar): in this case, via the analogy between the abstract poset structure (the partial ordering), used to define the theory, and the physical causality which motivates the theory. There is no more than an analogy here, because there really are no temporal relations; furthermore, one does not (need to) take a limit, in order to understand the theory. Physical insight can already be gained simply from consideration of the analogy.

Another example is loop quantum gravity, in its various versions (for an enlightening review, see Oriti (2014: §3.1)). In the original formulation, the Hilbert space of the theory was associated with graphs (which were discretisations of a spacetime): graphs were assigned elements of the group SU(2), and the wave-functions of the theory were functions of these group elements on graphs. Once again, we have here a case of analogy, as in (Similar): SU(2) is a group familiar from quantum mechanics, where it represents angular momentum and spin, even if in loop quantum gravity it receives a different physical interpretation. In fact, we recognise here not only the tool (Similar), but also (Internal) in §4.1: because once physicists realise that the theory has the mathematical structure of a quantum theory, then the different elements can be understood, on analogy with quantum mechanics, from the quantum-mechanical structure itself. For instance, once the wave functions have been set up, it is clear that certain operators which act on it can be interpreted as physical quantities, and among them there is a distinguished operator which plays the role of the energy.

In later formulations of loop quantum gravity, the so-called 'spin foams', the spacetime picture of the graphs (which were originally embedded in a spatial manifold) is dropped. One is left with a "theory of discrete combinational structures, the graphs, labelled by algebraic data only… one is really left with no reference to any underlying continuum spacetime, which has instead to be reconstructed from the combinatorial and algebraic data alone" (Oriti (2014: §3.1)).

One might object here that, while the theory contains no explicit reference to spacetime, it implicitly does so, for it was originally derived from a spacetime structure. But this is misunderstanding what is going on: it is not the case that the theory is 'derived' from spacetime structure; rather, the fundamental ontology of the theory now consists of algebraic and combinatorial data, from which the geometrical information, the spacetime, is *reconstructed* in a certain limit. It is the discrete structure that now may lead to spacetime and better intelligibility of the theory, and not the other way around. Notice that intelligibility is already present at the level of the underlying ontology, without the need for taking any limits. Indeed, we are here in the realm of (Similar) and (Internal) in §4.1, where physicists understand quantities through analogies, similarities, and structural arguments, precisely as discussed in the previous paragraph. For more details, as well as for a discussion of the requirement that a theory without a spacetime should have "no underlying spacetime", see De Haro and De Regt (2018).

This line of thought becomes more explicit in another quantum theory of gravity, called group field theory. Group field theory takes inspiration from the physical picture and methods of spin foams, but carries the algebraic (group) structure further. The theory now postulates that the basic fields take complex values on a group manifold, without a spacetime, and constructs an action out of such fields. Feynman diagram techniques for the group fields can be used, and techniques from quantum field theory can be borrowed, such as renormalisability analysis, invariance under U($N$) transformations, tensorial methods, etc. The interpretations of all these methods are now rather different from ordinary field theories which are based on a spacetime (in fact, this case is very similar to the random matrix models discussed in the previous Section). Nevertheless, the fact that this is 'just' a field theory gives physicists confidence that they understand what they are doing: after all, any quantum field theory has basically the same mathematical structure: a set of states and a set of operators, with eigenvalues of operators and expectation values being the quantities which can be compared to experimental outcomes. The same is true in loop quantum gravity and in group field theory. Thus, even though no effective spacetime needs to be derived in group field theory, physicists can rather straightforwardly make physical predictions, with ordinary methods from quantum field theory.

Note that the examples discussed in this section do not involve direct spatiotemporal visualization of reality. While some of the models discussed above employ certain forms of visualization (e.g. the Feynman diagram expansion of group field theory), this visualization need not have a spatiotemporal interpretation: it is simply a bookkeeping device. Thus, even if direct spatio-temporal visualization of reality is impossible, indirect forms of visualization may still be used to achieve understanding: visualization-as-bookkeeping turns out to be a further tool which is available in theories which have no spacetime at all.

### 4.4. Interpretation and the intelligibility of theories without a spacetime

The tools for understanding (Similar), (Internal), (Approximation), presented in §4.1 and applied in §4.2-4.3, are based on the notion of intelligibility as a process of *interpretation of elements* in a theory. The issue of interpretation indeed becomes particularly pressing in theories in which visualizability loses its power. As we argued in §2.1, interpretation typically leads to understanding, and is in fact a precondition for it. Without recognising mathematical structures in a theory, and providing them with a physical interpretation – through analogy, similarity, internal comparison, etc. – the theory cannot be rendered intelligible in the sense of $CIT_1$. Moreover, the tools for interpreting the theory are also the tools for rendering it intelligible (for a defence and further development of this thesis, see De Haro and De Regt 2018).

So how does this work for the case of theories without a spacetime? Does interpretation of these theories with the help of the tools (Similar), (Internal), and (Approximation) lead to intelligibility as defined in §2.1? To answer this question we should examine whether interpretation via these tools allows us to recognise qualitative consequences without exact calculation, as required by criterion $CIT_1$. For the case of (Approximation) this is obvious, since in all our examples the application of (Approximation) led to the appearance of effective spacetimes, which can be visualized. Hence, in these cases we can still use visualization as a tool to render the theory intelligible in accordance with $CIT_1$

But we have to explain how the kinds of interpretations that arise from (Similar) and (Internal) satisfy the criterion $CIT_1$. The connection is rather straightforward: an interpretation of terms in a physical theory usually enables physicists to think of the terms in the theory in specific ways. In the theories of quantum gravity that we have studied, there are no experiments yet, and the interpretations issuing from (Similar) and (Internal) are all structural: physicists interpret a *mathematical structure* or term in a model of the theory. The advantage of this kind of interpretations is that physicists are now able to think of a specific mathematical organising principle (an algebraic structure such as a group, or a Taylor series, etc.) in physically meaningful terms.

Understanding the theory now involves a *reversal* of the above process: once the connection from structure to interpretation is made (see De Haro (2016: §1.1.2), which describes interpretation in terms of appropriate maps), the physical interpretation of the structure can be used to predict mathematical outcomes, at least within certain approximations and context (since the maps involved are not invertible in the mathematical sense), without the need to do explicit calculations. Thus, the kind of understanding which physicists attain in this way is precisely $CIT_1$, in agreement with our analysis in Section 2.

## 5. Summary and Conclusion

In this paper our central question (BQ) has been whether theories without a spacetime can be rendered intelligible, so that they may provide scientific understanding. Adopting the contextual theory of scientific understanding, we argued in Section 2 that scientific understanding requires skills to work with the relevant theories. Visualization skills are often used to achieve understanding, but we argued that visualization is merely a 'contingently dominant' tool: although valued and employed by many scientists, it is not a necessary condition for intelligibility. We also invalidated metaphysical apriorism, a metaphysical position that is sometimes invoked to argue for the unintelligibility of theories without a spacetime. Thus, our answer to (BQ) is affirmative: on both epistemological and metaphysical grounds, it is possible to have understanding of theories without a spacetime. From an epistemological perspective, visualization and spacetime turn out to be valuable but not essential for intelligibility. And neither is spacetime a necessary part of the basic ontology of a theory, if we are to keep ontology open to whichever revisions our best scientific theories demand: as anyone, of course, should.

In Section 3, we analysed two discussions in philosophy of physics that relate to (BQ), namely the debate between Maudlin and Earman about Hamiltonian general relativity, and Lam and Esfeld's contribution to a special issue on emergence of spacetime. We saw that the stakes are high, for the authors do not hesitate to qualify the views that they oppose with terms such as 'absurdity', 'nonsense', and even 'crazy'. We argued that, in order to understand this polemic as a rational debate over the merits of a scientific theory, rather than as a matter of rhetoric and emotions (as some of the wording might suggest it to be), it should be construed as a debate on scientific understanding: what is at stake here is what counts as an *intelligible* theory. This underlines the importance of developing a theory of scientific understanding for theories without a spacetime.

Our analysis of these debates revealed two basic philosophical positions on the matter of understanding, which we have called 'intuitionism' and 'formalism'. These two camps differ in what they consider to be the essential elements of scientific understanding. For intuitionists, visualization in spacetime is essential for attaining understanding. Formalists, by contrast, are happy to give up visualization and other basic intuitions, if the physics so tells them, as long as there is a coherent physical picture—without an a priori constraint that this should be a spacetime picture. What, to one camp, appears as an abyss below which lies absurdity, is for the other camp a series of steps leading to understanding. The difference lies in what scientific understanding is and what it implies.

The contextual theory of understanding does not deny the value of spacetime and visualization as tools for understanding in particular contexts; on the contrary, it acknowledges that spatiotemporal visualization is a 'contingently dominant tool', and this explains the intuitionists' preference for it. It denies, however, that visualization in spacetime is a necessary condition for intelligibility and understanding. Our analysis of scientific understanding as a context-dependent notion aspires to be *descriptively accurate* of scientific practice, but this has the *normative implication* that Maudlin, Esfeld and Lam are wrong in their categorical rejection of theories without a spacetime as being unintelligible and-or metaphysically unacceptable. To be sure, there is nothing against using visualization as a tool for understanding (indeed, this is the motivation for constructing theories with effective spacetimes), but if this fails one should look for alternative tools for rendering the theories in question intelligible.

Having cleared the way for a constructive approach to the intelligibility of theories without a spacetime, we discussed three tools adequate for the task, that is, three procedures for interpreting the terms of a theory: (Similar), (Internal), and (Approximation). While the third tool may be used to construct effective spacetimes, and is thus related to visualization, the first two are tools for gaining scientific understanding without spacetime independently of visualization. The three tools are mainly aimed at interpreting terms in a theory, but we also argued that the result of using these tools conforms to $CIT_1$: interpreting a theory, using these tools, enables physicists to recognize qualitatively the characteristic consequences of models, without performing exact calculations. This is because an interpretation results in a specific organisation of the terms in a model; and that organisation allows for predictions without resorting to explicit calculation, which is what $CIT_1$ requires.

We studied two kinds of examples in support of our thesis: examples with and without effective spacetime (and thus visualization). In the case of theories in which an effective spacetime appears, this effective spacetime can appear independently of whether the theory has a spacetime in its basic ontology or not. This effective spacetime, which appears and is recognised as such through (Approximation), provides understanding of the theory in the usual way of visualization. The examples in which visualization and effective spacetimes are involved, thus illustrate visualization's dominance: most examples in quantum gravity indeed fall in this class: visualization is one of the most common tools. The contingency of visualization, on the other hand, is illustrated by the cases in which effective spacetime is absent, and not needed. In the latter cases, (Similar) and (Internal) can be used to gain understanding of a theory. This is achieved through the recognition of mathematical structures and of their physical roles (either through comparison to other theories, or internally). Such recognition of structures, and their physical interpretation through (Similar)-(Internal), leads to the identification of organising principles in the models, which by themselves are powerful enough so that physicists can use them to make predictions and take qualitative consequences of the theory, without doing explicit calculations. Thus, fulfilment of $CIT_1$ shows that even theories without an effective spacetime can be rendered intelligible. We conclude that scientific understanding does not require the presence of a spacetime of any sort.

**Acknowledgements**

We thank Jeremy Butterfield, Silvia De Bianchi, Carl Hoefer, and Daniele Oriti helpful discussions. We also thank the members of the research group on Philosophy of Science and Technology of VU University of Amsterdam for a discussion of the manuscript. SDH thanks audiences at the Ninth Workshop on the Philosophy of Information in Brussels, and at the Department of Philosophy of the Universitat Autònoma de Barcelona. SDH was supported by the Tarner scholarship in Philosophy of Science and History of Ideas, held at Trinity College, Cambridge.

**References**

Banks, T., Fischler, W., Shenker, S.H. and Susskind, L. (1997). 'M Theory as a Matrix Model: A Conjecture'. *Physical Review*, D55, p. 5112. doi:10.1103/PhysRevD.55.5112 [hep-th/9610043].

Baumberger, Christoph, Claus Beisbart, and Georg Brun. (2017). 'What is understanding? An overview of recent debates in epistemology and philosophy of science'. In: Grimm, Baumberger and Ammon (2017), pp. 1-34.

Beller, Mara. (1999). *Quantum Dialogue: The Making of a Revolution*, *Science and its conceptual foundations*. Chicago: University of Chicago Press.

Brézin, E., Itzykson, C., Parisi, G., and Zuber, J.B. (1978). 'Planar Diagrams'. *Communications in Mathematical Physics*, 59, pp. 35-51.

Butterfield, Jeremy N. (2011). 'Emergence, reduction and supervenience: a varied landscape', *Foundations of Physics*, 41 (6), pp. 920-959.

Butterfield, Jeremy N. and Bouatta, Nazim (2015). 'On Emergence in Gauge Theories at the 't Hooft Limit'. *European Journal for Philosophy of Science*, 5 (1), pp. 55-87.

Cushing, James T. (1994). *Quantum mechanics: Historical Contingency and the Copenhagen Hegemony*, *Science and its conceptual foundations*. Chicago: University of Chicago Press.

Dardashti, Radin, Karim PY Thébault, and Eric Winsberg (2015). 'Confirmation via analogue

simulation: what dumb holes could tell us about gravity'. *British Journal for the Philosophy of Science* 68 (1), pp. 55-89.

De Haro, Sebastian (2015). 'Dualities and emergent gravity: Gauge/gravity duality'. *Studies in History and Philosophy of Modern Physics*, 59, 2017, pp. 109-125. doi: 10.1016/j.shpsb.2015.08.004.

De Haro, Sebastian (2016). 'Spacetime and Physical Equivalence'. Forthcoming in *Space and Time after Quantum Gravity*, Huggett, N. and Wüuthrich, C. (Eds.). http://philsci-archive.pitt.edu/13243.

De Haro, Sebastian (2017). 'Towards a Theory of Emergence for the Physical Sciences'. In preparation.

De Haro, Sebastian, Teh, Nicholas, Butterfield, Jeremy N. (2015). 'Comparing dualities and gauge symmetries'. *Studies in History and Philosophy of Modern Physics*, 59, 2017, pp. 68-80. https://doi.org/10.1016/j.shpsb.2016.03.001

De Haro, Sebastian and Butterfield, Jeremy N. (2017). 'A Schema for Duality, Illustrated by Bosonization', to appear in a volume dedicated to the centenary of Hilbert's work on the foundations of Mathematics and physics: *Foundations of Mathematics and Physics one century after Hilbert*. Kouneiher, J. (Ed.). Collection Mathematical Physics, Springer. http://philsci-archive.pitt.edu/13229.

De Haro, Sebastian and De Regt, Henk W. (2018). 'Interpreting Theories without Spacetime'. Forthcoming in *European Journal for Philosophy of Science*, https://doi.org/10.1007/s13194-018-0204-x.

De Regt, Henk W. (2009). 'The epistemic value of understanding', *Philosophy of Science* **76**, pp. 585-597.

De Regt, Henk W. (2013). 'Understanding and explanation: Living apart together?' *Studies in History and Philosophy of Science* **44**, pp. 505-510.

De Regt, Henk W. (2014). 'Visualization as a tool for understanding', *Perspectives on Science* **22**, pp. 377-396.

De Regt, Henk W. (2017), *Understanding Scientific Understanding*. New York: Oxford University Press.

De Regt, Henk W. and Dennis Dieks (2005). 'A contextual approach to scientific understanding', *Synthese* 144, pp. 137-170.

De Regt, Henk W., Sabina Leonelli & Kai Eigner (eds.) (2009). *Scientific Understanding: Philosophical Perspectives*. Pittsburgh: University of Pittsburgh Press.

Dijkgraaf, R., and Vafa, C. (2002). 'Matrix Models, Topological Strings, and Supersymmetric Gauge Theories'. *Nuclear Physics*, B 644, p. 3. doi:10.1016/S0550-3213(02)00766-6. [hep-th/0206255].

Earman, John (2002). 'Thoroughly Modern McTaggart. Or what McTaggart would have said if he had learned General Relativity Theory', *Philosopher's Imprint* 2, pp. 1-28.

Grimm, S.R., C. Baumberger and S. Ammon (eds.) (2017), *Explaining Understanding. New Perspectives from Epistemology and Philosophy of Science*. New York: Routledge.

Gryb, S., and Thébault, K. P. (2014). 'Time remains'. *The British Journal for the Philosophy of Science*, 67(3), 2015, pp. 663-705.

Huggett, Nick, Vistarini, Tiziana, and Wüthrich, Christian (2012). 'Time in quantum gravity'.

arXiv preprint arXiv:1207.1635. Forthcoming in Adrian Bardon and Heather Dyke (eds.), *The Blackwell Companion to the Philosophy of Time*.

Huggett, Nick, and Wüthrich, Christian (2013). 'The emergence of spacetime in quantum theories of gravity'. *Studies in the History and Philosophy of Modern Physics*, 44(3), pp. 273-275.

Kaiser, David. (2005). *Drawing Theories Apart. The Dispersion of Feynman Diagrams in Postwar Physics*. Chicago: The University of Chicago Press.

Lam, Vincent and Esfeld, Michael (2013). 'A dilemma for the emergence of spacetime in canonical quantum gravity'. *Studies in History and Philosophy of Modern Physics,* 44, pp. 286-293.

Lipton, Peter. (2004). *Inference to the Best Explanation*. 2nd ed, *International library of philosophy*. London: Routledge.

Mariño, Marcos (2004). 'Les Houches Lectures on Matrix Models and Topological Strings'. arXiv preprint hep-th/0410165.

Maudlin, Tim (2002). 'Thoroughly muddled McTaggart: Or, how to abuse gauge freedom to create metaphysical monostrosities'. *Philosopher's Imprint* 2 (4), pp. 1-23.

Newman, M. (2012). 'An inferential model of scientific understanding'. *International Studies in the Philosophy of Science,* 26 (1), 1-26.

Morgan, Mary, and Margaret Morrison (eds.) (1999). *Models as Mediators: Perspectives on Natural and Social Science*. Cambridge: Cambridge University Press.

Morrison, Margaret. (2015). *Reconstructing Reality: Models, Mathematics, and Simulations*. New York: Oxford University Press.

Möβner, Nicola. (2015). 'Visual Information and Scientific Understanding'. *Axiomathes*, 25, pp. 167-179.

Oriti, Daniele (2014). 'Disappearance and emergence of space and time in quantum gravity'. *Studies in History and Philosophy of Modern Physics*, 46, pp. 186-199.

Pitts, J. Brian (2014). 'Change in Hamiltonian general relativity from the lack of a time-like Killing vector field'. *Studies in History and Philosophy of Modern Physics*, 47, pp. 68-89.

Schrödinger, Erwin. (1928). *Collected Papers on Wave Mechanics*. London: Blackie.

't Hooft, Gerard (1974). 'A Planar Diagram Theory for the Strong Interactions'. *Nuclear Physics*, B72, pp. 461-473.

Wilkenfeld, D. A. (2013). 'Understanding as representation manipulability'. *Synthese,* 190: 997–1016.

Wüthrich, Christian and Callender, Craig (2016). 'What Becomes of a Causal Set?' *The British Journal for the Philosophy of Science*, 68 (3), pp. 907-925.